\documentclass[a4paper,10pt,twocolumn,twoside]{article}
\usepackage{amssymb}
\usepackage{amsfonts}
\usepackage{latexsym}
\usepackage{layout}
\usepackage{amsmath}
\usepackage{graphicx}
\usepackage[dvips]{color}

\begin{document}

\title{Variation of incoming solar radiation flux during a partial eclipse episode: an improved model simulation}

\author{Boyan Petkov $^{1,2}$          \and
        Claudio Tomasi $^{1}$          \and
        Vito Vitale $^{1}$             \and
        Christian Lanconelli $^{1}$    \and
        Mauro Mazzola $^{1}$\\
\\$^{1}$Institute of Atmospheric Sciences and Climate (ISAC) of the
Italian
\\National Research Council (CNR), Via Gobetti 101,
I-40129 Bologna, Italy\\
\\
$^{2}$International Centre for Theoretical Physics (ICTP), \\TRIL
Programme, Strada Costiera 11, I-34014 Trieste, Italy}

\date{\textit{\today}}

\maketitle

\begin{abstract}\small Model simulations of solar irradiance reaching the
Earth's surface during a solar eclipse constitute a useful tool for
studying the impact of this phenomenon on the radiance propagation
through the atmosphere. A simple approach to extend the use of an
algorithm already adopted for evaluating the variations in the
extraterrestrial solar radiation during a total eclipse is proposed
for a partial eclipse case. The application is based on the
assessment of the distance between the apparent solar and lunar disk
centers on the celestial hemisphere, using the local circumstances
and the ratio between the Sun and Moon radii as input parameters. It
was found that during the eclipse of March 29, 2006, the present
approach led to an estimate of the surface UV solar irradiance trend
differing by no more than  $\pm5\%$ from the corresponding trend
observed at Bologna (Italy).
\end{abstract}

\section{\large Introduction}
\label{intro} A solar eclipse is a good opportunity for studying the
interaction between solar radiation and the Earth's atmosphere,
\cite{[RefPenaloza],RefAbbott,RefAnderson,RefZerefos,RefKazadsis},
and for testing radiative transfer models \cite{RefEmde}. Together
with ground-based measurements, modeling of the incoming solar
irradiance at the surface during an eclipse episode can provide
useful information for those investigating such interaction
processes \cite{RefKazadsis}.

M\"{o}llman and Vollmer \cite{RefMollmann} proposed an approach
aimed to evaluate the broadband solar irradiance during a solar
eclipse, in which the limb darkening effect was neglected. An
accurate procedure for calculating the spectral variations in the
extra-terrestrial solar radiation occurring during a total eclipse
was developed by Koepke et al. \cite{RefKoepke}. The approach takes
into account the limb darkening and determines the radiance changes
in terms of normalized spectral irradiance $I_{norm,\lambda}$
expressed as the ratio between the radiation coming from the part of
the solar disk uncovered by the Moon and the radiation emitted by
the entire solar disk. The parameter $I_{norm,\lambda}$
characterizes the variations in solar irradiance at the top of the
atmosphere during an eclipse episode and, therefore, it can be used
as a correction factor for evaluating the extra-terrestrial
irradiance in the radiative transfer models. The main parameter
describing the eclipse geometry in the method is the distance
between the apparent solar and lunar disks, which was evaluated in
the total eclipse case. The present study aims to extend the
approach developed by Koepke et al. \cite{RefKoepke} to a partial
eclipse case, assessing the length of segment defined by the Sun and
Moon centers as a function of the ratio between the corresponding
radii and local eclipse circumstances.

\section{\large Assessment of the solar-moon apparent distance during a partial solar eclipse}
\label{sec1} In case of total eclipse, the apparent path of the
lunar disk center on the celestial hemisphere, crosses the
corresponding center of the solar disk, and parameter
$I_{norm,\lambda}$ can be determined \cite{RefKoepke} as a function
of the distance $X$ between the Sun and Moon centers, which is
assumed to follow a linear time-dependent trend for a constant
velocity $v$ of the Moon with respect to the Sun. However, in the
case of a partial solar eclipse, the Moon disk centre does not cross
the corresponding solar center \cite{RefDuffettSmith}, as is shown
in Fig. \ref{fig1} and, therefore, parameter $X$ does not vary
linearly in time. The graph represents the relative movements of the
apparent solar and lunar disks of radii $R_{S}$ and $R_{M}$,
respectively, during an eclipse episode starting at time $t_{o}$,
reaching its maximum at time $t_{m}$ and finishing at time $t_{e}$.
All the distances shown in Fig. \ref{fig1} are considered to be
normalized to parameter $R_{S}$, so that the apparent Sun disk
radius is considered as unit ($R_{S}=1$). In addition, the solar
disk has been assumed to have a fixed position. Due to the specific
geometry of the problem, the time-interval between $t_{o}$ and
$t_{m}$ is usually slightly different from the corresponding
time-interval between $t_{m}$ and $t_{e}$. This is why the eclipse
episode is arbitrarily subdivided into two parts, the former of
which is taken between times $t_{o}$ and $t_{m}$ and  labeled with
apex ($'$), and the latter between times $t_{m}$ and $t_{e}$ with
apex ($''$). According to Fig. \ref{fig1}, parameter $X'$ can
therefore be expressed as:

\begin{figure} [t]
\includegraphics[width=0.48\textwidth]{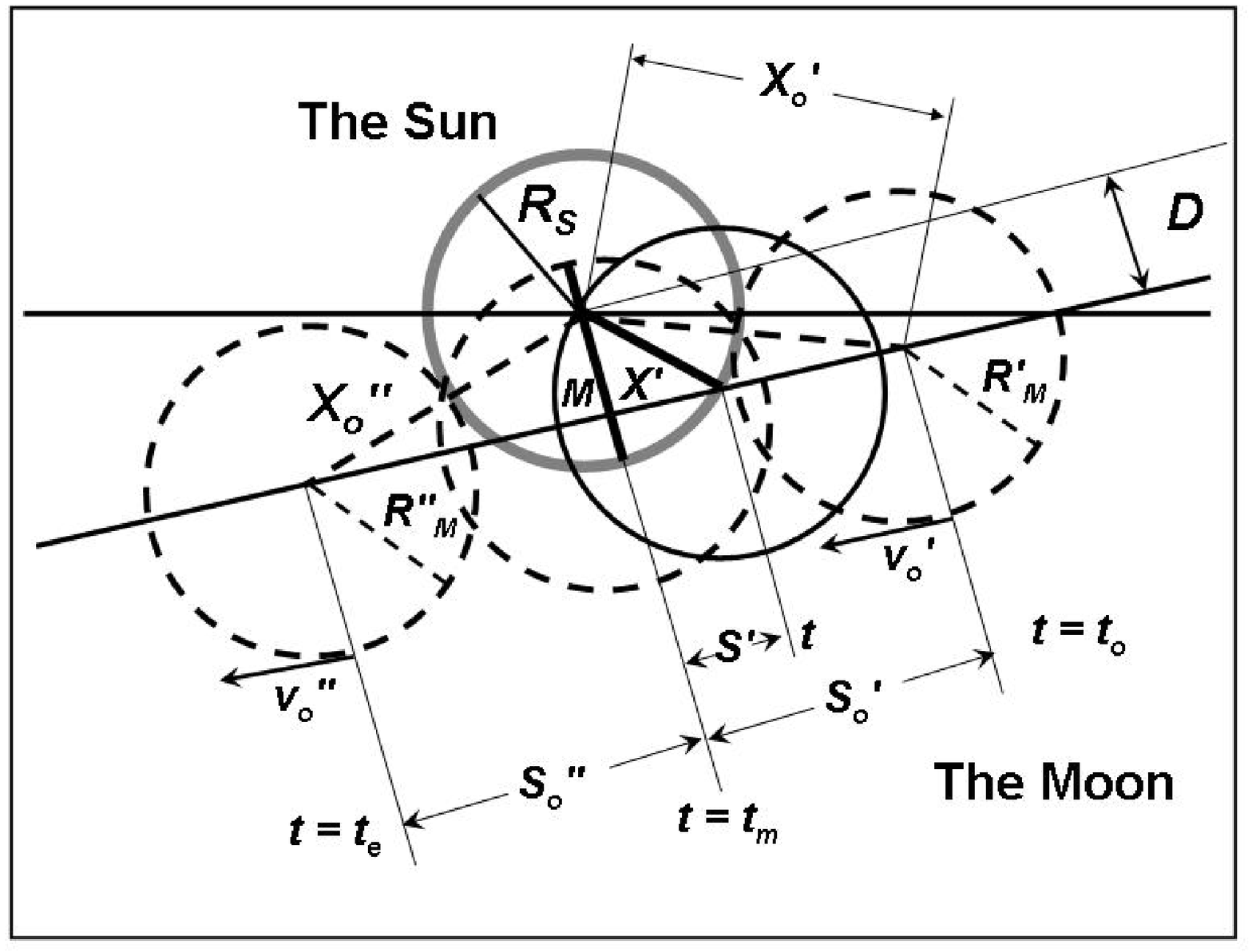}
\caption {\scriptsize Apparent movement of the Moon disk of radius
$R_{M}$, labeled with apex $(')$ for the first phase of the eclipse
and apex $('')$ for the second, with respect to the Sun disk of
radius $R_{S}$, during a partial solar eclipse episode. The dashed
circles represent the position of the Moon at times relative to
(from right to left): (i) first contact, (ii) maximum eclipse, and
(iii) fourth contact. The distance $X$ between solar and lunar disk
centers is similarly labeled with apices $(')$ and $('')$ for two
phases of the eclipse, respectively. $D$ is the distance between the
Sun and Moon disk centers at the maximum eclipse time $t_{m}$; $M$
is the part of the solar disk radius covered by the Moon at time
$t_{m}$; and $S_{o}$ (and $S$ at time $t$) are the apparent paths of
the Moon disk center during the eclipse, labeled with apices $(')$
and $('')$, respectively, as for the above-mentioned parameters.}
\label{fig1}
\end{figure}

\begin{equation}\label{eq1}
X'\,=\,\sqrt{D^2+S'^{2}} \, .
\end{equation}
Parameter $D$ is the distance between the centers of apparent Sun
and Moon disks at the maximum eclipse time and can be determined in
terms of the following equation,
\begin{equation}\label{eq2}
D\,=\,R_{S}+R^{0}_{M}-M \,=\, 1+R^{0}_{M}-M \,  ,
\end{equation}
where $R^{0}_{M}$ is the Moon radius at time $t_{m}$. Segment $M$,
which represents the part of solar radius $R_{S}$ covered by the
Moon at the maximum eclipse time $t_{m}$, can be expressed in terms
of the eclipse magnitude $M_{g}$ for the considered site in the
following form:
\begin{equation}\label{eq3}
M\,=\,2R_{S}M_{g}\,=\,2M_{g}\, .
\end{equation}
At the eclipse start time $t_{o}$, parameter $X'_{o}$ is equal to
$R_{S}+R'_{M}$ and, hence, during the period $[t_{o},t_{m}]$, the
Moon disk centre describes the path $S_{o}'$ shown in Fig.
\ref{fig1}, whose length is given by
\begin{multline}\label{eq4}
S_{0}'=\sqrt{(R_{S}+R'_{M})^{2}-D^{2}}=\\=\sqrt{(1+R'_{M})^{2}-D^{2}}
\, .
\end{multline}
Assuming nearly constant velocity $v_{0}'$ of the Moon with respect
to the Sun during the first phase of the eclipse, it can be stated
that
\begin{equation}\label{eq5}
v_{0}'\,=\, \frac{S_{0}'}{t_{m}-t_{0}} \, .
\end{equation}
Therefore, parameter $S'$ can be defined at each time
$t\in[t_{0},t_{m}]$ in the explicit form,
\begin{equation}\label{eq6}
S'\,=\, S_{0}'-v_{0}'(t-t_{0})\, .
\end{equation}
Thus, the distance $X'$ can be defined using Eqs (\ref{eq1}) --
(\ref{eq6}). Similarly, the corresponding parameter $S''$ can be
expressed as equal to the difference:
\begin{equation}\label{eq7}
S''\,=\, S''_{0}-v_{0}''(t_{e}-t) \, ,
\end{equation}
where $t\in(t_{m},t_{e}]$ and parameters $S''_{0}$ and $v_{0}''$ are
defined according to the following pair of equations:
\begin{equation}\label{eq8}
S_{0}'' \,=\, \sqrt{(1+R''_{M})^{2}-D^{2}}
\end{equation}
and
\begin{equation}\label{eq9}
v_{0}'' \,=\, \frac{S_{0}''}{t_{e}-t_{m}}\,\, ,
\end{equation}
assuming again that velocity $v_{0}''$ is nearly constant. Finally,
the distance $X''$ between the apparent solar and lunar disk centers
during the second phase of the eclipse is defined according to Eq
(\ref{eq1}), as
\begin{equation}\label{eq10}
X'' \,=\, \sqrt{D^2+S''^{2}} \, .
\end{equation}

\begin{table}[t]
\caption{\label{tab1} \scriptsize Parameters of the solar eclipse of
29 March 2006 estimated by Espenak and Anderson \cite{RefEspenak}
for Bologna, Italy. The values of ratio $R_{M}$/${R_{S}}$ are taken
from physical ephemeris of the umbral shadow (Table 4 in
\cite{RefEspenak}).}\scriptsize
\begin{tabular}{cccc}
\hline\noalign{\smallskip}
Contact &  Time &  $R_{M}$/${R_{S}}$  &    Magnitude    \\
        & (UTC) & $({R_{S}}\,=\, 1)$  &     $M_{g}$     \\
\noalign{\smallskip}\hline\noalign{\smallskip}
I & 09:33:01 ($t_{o}$)& 1.0500  ($R'_{M}$)& ---\\
Maximum & 10:38:03 ($t_{m}$) & 1.0509 ($R^{o}_{M}$)& 0.532 \\
eclipse &  &  &  \\IV & 11:44:11 ($t_{e}$)& 1.0375  ($R''_{M}$)& --- \\
\noalign{\smallskip}\hline
\end{tabular}
\end{table}

Equations (\ref{eq1}) -- (\ref{eq10}) provide an algorithm suitable
for determining the time-patterns of parameter $X$, denoted as $X'$
during the first half of the event and as $X''$ during the second
half, knowing the local circumstance times, eclipse magnitude
$M_{g}$ and behavior of the ratio $R_{M}/{R_{S}}$ during the
eclipse. In fact, the present procedure takes into account the ratio
$R_{M}/{R_{S}}$ only at the start time $t_{o}$ of the event, at the
maximum eclipse time $t_{m}$, and at the final time $t_{e}$, which
correspond to parameters $R'_{M}$, $R^{o}_{M}$ and $R''_{M}$,
respectively, assuming $R_{S}=1$ as established above. Since the
variations in ratio $R_{M}/{R_{S}}$ usually do not exceed 1-2\%, the
error made in estimating parameter $X$ when considering such a ratio
at three fixed times only, will be around 1\%. In case of total
eclipse, parameter $D$ is equal to zero and the above approach gives
a linear trend of distance $X$, as assumed by Koepke et al.
\cite{RefKoepke} It is also worth noting that Koepke et al.
\cite{RefKoepke} normalized the parameter $X$ by $R_{M}+{R_{S}}$ in
their procedure, while such a step was avoided in the present
procedure for sake of simplicity. The above algorithm is similar to
that proposed by M\"{o}llman and Vollmer \cite{RefMollmann} to
evaluate the illuminance during a solar eclipse. The main difference
is the subdivision of the event into two periods, before the maximum
phase and after that, while M\"{o}llman and Vollmer
\cite{RefMollmann} assumed a symmetry of the eclipse episode
evolutionary features with respect to the maximum. On the other hand
they assessed the broad-band irradiance and, hence, the limb
darkening was not taken into account. Conversely, the Koepke et al.
\cite{RefKoepke} method considers this effect: based on these
concepts, the present algorithm is suitable to estimate the spectral
variations of surface solar irradiance during a partial solar
eclipse.

\begin{figure}[h]
\includegraphics[width=0.48\textwidth]{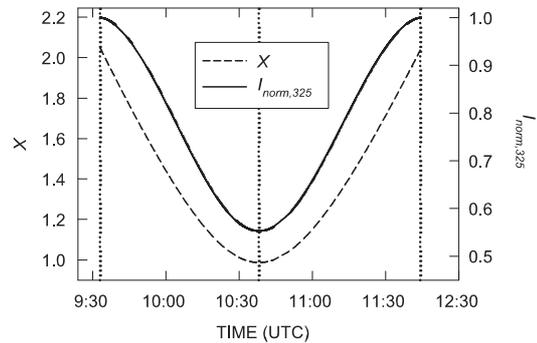}
\caption{\scriptsize Time-patterns of parameters $X$ and
$I_{norm,325}$ during the partial solar eclipse of 29 March 2006
evaluated for Bologna (Italy): $X$ is the distance between the Moon
and Sun disk centers; $I_{norm,325}$ is the normalized solar
irradiance at the 325 nm wavelength, given by the ratio between the
radiation coming from the uncovered part of the solar disk and that
coming from the whole disk. The contact times and maximum eclipse
time (see Table 1) are indicated by vertical dotted lines.}
\label{fig2}
\end{figure}

\section{\large Validation}
\label{sec2} To illustrate the outputs given by the present
algorithm, the global solar UV irradiance time-patterns observed
during the partial solar eclipse of March 29, 2006 at Bologna, Italy
($44^{o}31'$N, $11^{o}20'$E), were evaluated and compared with the
ground-based measurements \cite{RefPetkov1} performed with the
narrow-band filter radiometer UV-RAD \cite{RefPetkov2}. Table
\ref{tab1} presents the input parameters, as provided by Espenac and
Anderson \cite{RefEspenak}. Figure \ref{fig2} shows the
time-patterns of normalized solar irradiance $I_{norm,325}$ at the
325 nm wavelength, as calculated following the Koepke et al.
\cite{RefKoepke} procedure, together with those of distance X
evaluated through Eqs (\ref{eq1}) -- (\ref{eq10}). Parameter
$I_{norm,325}$ defines the relative variations of the radiance
entering the atmosphere at the 325 nm wavelength, occurring at
Bologna during the event. It is suitable for use as a correction
factor for extra-terrestrial solar radiation in a radiative transfer
model used to evaluate the surface solar irradiance. To perform such
an evaluation, the widely used TUV model \cite{RefMadronich} was
applied for the 325 nm spectral component of solar irradiance,
keeping constant the daily mean value of columnar ozone amount equal
to 390 DU, as determined by UV-RAD on March 29, 2006. Since the
eclipse day was characterized by stable clear-sky conditions, cloud
effects were not considered in the present assessment. On the basis
of ground-level visibility evaluations made on the eclipse day
\cite{RefPetkov1}, an aerosol volume extinction coefficient was
assumed at the 550 nm wavelength varying between less than 0.04 and
0.05 km$^{-1}$. Consequently, assuming a particle scale height close
to 1.25 km according to the Penndorf \cite{RefPenndorf} evaluations
(confirmed by the estimates of 1.0 - 1.5 km made by Tomasi
\cite{RefTomasi} in the Po Valley on early spring days), aerosol
optical depth (AOD) at visible wavelengths can be evaluated to range
between 0.04 and 0.07, for meteorological and atmospheric
transparency conditions similar to those of the eclipse day. Thus,
an AOD value equal to 0.05 was given as input to the code, together
with the Elterman \cite{RefElterman} vertical distribution of the
volume extinction coefficient of aerosol particles. The winter
Mid-latitude temperature profile determined by Anderson et al.
\cite{RefAndersonetal} was used in the code calculations for surface
albedo equal to 0.08, leading to the best agreement between
evaluated and measured irradiances during the part of the day in
which perturbations due to the eclipse were not observed.

\begin{figure}[h]
\includegraphics[width=0.48\textwidth]{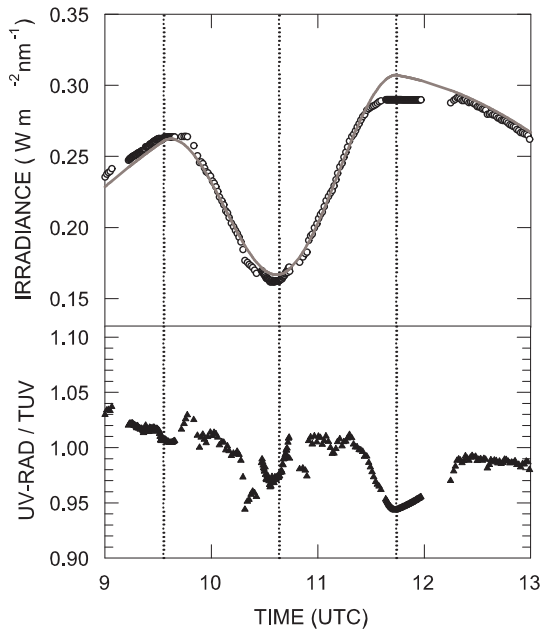}
\caption{\scriptsize Upper part: time-patterns of surface UV solar
irradiance at the 325 nm wavelength, measured during the partial
solar eclipse of 26 March 2006 at Bologna (open circles). The thick
grey curve represents the corresponding model evaluations. Lower
part: time-patterns of the ratio between UV-RAD measured and
model-evaluated irradiances, respectively. The contact times and the
maximum eclipse time (see Table \ref{tab1}) are indicated by
vertical dotted lines.} \label{fig3}
\end{figure}

In the upper part of Fig. \ref{fig3}, the results of the simulation
are presented, showing a comparison between the theoretical and
observed variations in solar irradiance. The lower part of Fig.
\ref{fig3} presents the time-patterns of the ratio between measured
and calculated irradiance values, providing evidence of the good
agreement between the two variables. Except for the times of maximum
eclipse and last contact, when discrepancies of about -5\% were
found, all the other cases exhibit similar differences within
$\pm3\%$. The comparatively higher deviations between evaluated and
measured values at times $t_{m}$ and $t_{e}$ can be reasonably
attributed to the presence of sparse \textit{cumulus congestus}
clouds moving around the Sun on 29 March 2006, causing, together
with the eclipse event, appreciable changes in the surface solar
irradiance that cannot be represented with good precision by using a
monodimensional radiative transfer model.

\section{\large Conclusions}
\label{sec3} A simple algorithm has been proposed suitable for
extending the method of Koepke et al. \cite{RefKoepke}, developed to
correct the trend of extra-terrestrial solar radiation during a
total solar eclipse to a partial eclipse case. To do this, the
evolution of the distance between the apparent Moon and Sun disk
centers was evaluated as a function of local eclipse circumstances
and the ratio between the lunar and solar radii. The estimation
allows the calculation of the variations in the solar irradiance
entering the terrestrial atmosphere, where the surface irradiance
time-patterns are simulated using a radiative transfer model. It was
found that the procedure provides a satisfactory assessment of the
evolutionary patterns characterizing the UV solar irradiance during
the partial eclipse of March 29, 2006 computed by the TUV model
applied to the Bologna (Italy) site.

\textheight=320pt


\begin{thebibliography}{XXXXXX}
\scriptsize
\bibitem{[RefPenaloza]}
M. A. Penaloza-Murillo, Optical response of the atmosphere during
the Caribbean total solar eclipse of 26 February 1998 and of 3
February 1916 at Falcyn State, Venezuela.  Earth Moon Planets 91,
(2002), 125-159.

\bibitem{RefAbbott}
W. N. Abbott, On certain radiometric effects during the partial
eclipse of February 25, 1952.  Geofis. Pura Appl. (Milan), 39,
186-193 (1958).

\bibitem{RefAnderson}
R. C. Anderson, D. R. Keefer, and O. E. Myers, Atmospheric pressure
and temperature changes during the 7 March 1970 solar eclipse. Jour.
Atmos. Sci., 29, 583-587 (1972).

\bibitem{RefZerefos}
C. Zerefos, D. S. Balis, C. Meleti, A .F. Bais, K. Tourpali, K.
Kourtidis, K. Vanicek, F. Cappellani, U. Kaminski, T. Colombo, R.
St\"{u}bi, L. Manea, P. Formenti, M. O. Andreae, Changes in surface
solar UV irradiances and total ozone during the solar eclipse of
August 11, 1999. J. Geophys. Res., 105, 26463-26473 (2000).

\bibitem{RefKazadsis}
S. Kazadsis, A. Bais, M. Blumthaler, A. Webb, N. Kouremeti, R. Kift,
B. Schallhart, and A. Kazantzidis, Effects of total solar eclipse of
29 March 2006 on surface radiation. Atmos. Chem. Phys. Discuss., 7,
9235-9258, (2007).

\bibitem{RefEmde}
C. Emde, and B. Mayer, Simulation of solar radiation during a total
eclipse: a challenge for radiative transfer. Atmos. Chem. Phys., 7,
2259-2270 (2007).

\bibitem{RefMollmann}
K.-P. M\"{o}llmann and M. Vollmer, Measurements and predictions of
the illuminance during a solar eclipse. Eur. J. Phys., 27, 1299-1314
(2006)

\bibitem{RefKoepke}
P. Koepke, J. Reuder, and J. Schween, Spectral variation of the
solar radiation during an eclipse. Meteorol. Zeitschrift, 10,
179-186 (2001).

\bibitem{RefDuffettSmith}
P. Duffett-Smith, Astronomy with your personal computer, 2nd edn.
(Cambridge University Press, Cambridge, 1990), p. 179.

\bibitem{RefPetkov1}
B. Petkov, C. Tomasi, V. Vitale, A. di Sarra, P.
Bonasoni, C. Lanconelli, E. Benedetti, D. Sferlazzo, H. Diémoz, G.
Agnesod, R. Santaguida, "Ground-based observations of solar
radiation at three Italian sites, during the eclipse of 29 March,
2006: Signs of the environment impact on incoming global
irradiance," Atmos. Res. 96, 131-140 (2010).

\bibitem{RefPetkov2}
B. Petkov, V. Vitale, C. Tomasi, U. Bonaf\'{e}, S. Scaglione, D.
Flori, R. Santaguida, M. Gausa, G. Hansen, and T. Colombo,
Narrow-band filter radiometer for ground-based measurements of
global UV solar irradiance and total ozone. Appl. Opt. 45, 4383-4395
(2006).

\bibitem{RefEspenak}
F. Espenak and J. Anderson, Total Solar Eclipse of 2006 March 29,
NASA/TP-2004-212762, 2004 (availible in
http://eclipse.gsfc.nasa.gov/SEmono/TSE2006/TSE2006.\\html).

\bibitem{RefMadronich}
S. Madronich, UV radiation in the natural and perturbed atmosphere,
in "Environmental Effects of UV (Ultraviolet) Radiation" (M. Tevini,
ed), pp. 17-69. Lewis, Boca Raton (1993).

\bibitem{RefPenndorf}
R. Penndorf, The vertical distribution of Mie particles in the
troposphere. J. Meteorol. 11, 245-247 (1954).

\bibitem{RefTomasi}
C. Tomasi, Features of the scale height for particulate extinction
in hazy atmospheres. J. Appl. Meteor. 21, 931-944 (1982).

\bibitem{RefElterman}
L. Elterman, UV, visible, and IR attenuation for altitudes to 50 km.
Environmental Research Papers, 285, Report 68-0153, Air Force
Cambridge Research Laboratories (1968).

\bibitem{RefAndersonetal}
G. P. Anderson, S. A. Clough, F. X. Kneizys, J. H. Chetwynd, E. P.
Shettle, AFGL Atmospheric Constituent Profiles (0 - 120 km).
Environ. Res. Pap. 954, Opt. Phys. Div., Air Force Geophys. Lab.,
Hanscom Air Force Base, Mass. (1986).

\end{thebibliography}
\end{document}